\documentclass{jfm}
\usepackage{amsmath,amsfonts,amssymb,bm,upmath,graphicx}
\newcommand{\ud}{\mathrm{d}}

\title{Point-source scalar turbulence}
\author[A. Celani, M. Martins Afonso and A. Mazzino]{A\ls N\ls T\ls O\ls N\ls I\ls O\ns C\ls E\ls L\ls A\ls N\ls I$^1$,\ns M\ls A\ls R\ls C\ls O\ns M\ls A\ls R\ls T\ls I\ls N\ls S\ns A\ls F\ls O\ls N\ls S\ls O$^{2}$ and A\ls N\ls D\ls R\ls E\ls A\ns M\ls A\ls Z\ls Z\ls I\ls N\ls O$^3$}
  \affiliation{$^1$CNRS, INLN, 1361 Route~des~Lucioles, 06560~Valbonne, France\\[\affilskip]
               $^2$Department~of~Physics~of~Complex~Systems, The~Weizmann~Institute~of~Science, Rehovot~76100, Israel\\[\affilskip]
               $^3$Department~of~Physics~-~University~of~Genova \& CNISM and INFN~-~Genova~Section, via~Dodecaneso~33, 16146~Genova, Italy}
  \date{\today}

\usepackage{natbib}
\allowdisplaybreaks[1]

\begin{document}

\maketitle

\begin{abstract}
 The statistics of a passive scalar randomly emitted from a point source is investigated analytically. Our attention has been focused on the two-point equal-time scalar correlation function. The latter is indeed easily related to the spectrum, a statistical indicator widely used both in experiments and in numerical simulations. The only source of inhomogeneity/anisotropy is in the injection mechanism, the advecting velocity here being statistically homogeneous and isotropic. Our main results can be summarized as follows. 1) For a very large velocity integral scale, a pure scaling behaviour in the distance between the two points emerges only if their separation is much smaller than their distance from the point source. 2) The value we have found for the scaling exponent suggests the existence of a direct cascade, in spite of the fact that here the forcing integral scale is formally set to zero. 3) The combined effect of a finite inertial-range extension and of inhomogeneities causes the emergence of subleading anisotropic corrections to the leading isotropic term, that we have quantified and discussed.
\end{abstract}

\section{Introduction}

 Turbulent fields whose statistical properties are invariant under translation and rotation in space are particularly interesting to theorists. They represent a formally simple setup allowing to neglect additional `complications' introduced by the boundaries of the system or external driving mechanisms. However, in most situations of interest, such complications might play a crucial role in determining the statistical properties of the turbulent field. For the sake of example, this is the case in channel-flow turbulence, where statistical invariance under translation and rotation (i.e.~homogeneity and isotropy) is only restricted to a small region around the center of the channel, while it is totally lost close to the walls \cite[][]{TASBP99}.\\
 In the last few years, important achievements have been obtained, both for the method of analysis and the level of comprehension, in relation to the effects on the small-scale statistics of anisotropic large-scale contributions \cite[see][for a review]{BP05}. The situation is much less clear for what concerns the role of inhomogeneities in determining the statistics of small-scale turbulence. In particular, it is not clear if inhomogeneities, activated by boundaries and/or initial conditions, might disappear at small scales owing to cascade processes, which tend to eliminate a detailed memory of the large-scale dynamics. In terms of scaling invariance of small-scale statistics, it is unclear if the presence of inhomogeneities might lead to different scenarios with respect to the case where inhomogeneities are absent. Understanding the above aspects has relevant consequences in applications related, for example, to small-scale subgrid parameterizations. Indeed, modern approaches to closure problems commonly use scaling exponents as the basic ingredients to build subgrid scale models. The best examples are the fractal \cite[][]{SM97} and multifractal \cite[][]{BFP04} interpolation schemes. How to define scaling exponents in the presence of anisotropies became clear only very recently \cite[][]{BP05}. The situation is different in the presence of inhomogeneities, where, up to now, also basic questions related to the existence of scaling behaviour, and thus of scaling exponents, still have to be seriously addressed.\\
 The main aim of our paper is to give quantitative answers to the above questions. To do that, we will focus our attention on passive scalar turbulence, where the scalar is randomly emitted from a point source. The injection mechanism is thus intimately inhomogeneous and the question on how this inhomogeneity eventually reflects on the small-scale scalar statistics can be addressed. In order to carry out the study in analytical terms, we will assume a white-in-time, homogeneous, random process to model the velocity-field statistics \cite[][]{K68,K94}.\\
 The paper is organized as follows. In \S~\ref{sec:be} we formulate the problem in the context of the Kraichnan advection model with inhomogeneous forcing and we adopt simple mathematical techniques to obtain the equation for the two-point equal-time scalar correlation function, whose general solution is provided in \S~\ref{sec:gs} for a point-source emission. In \S~\ref{sec:lc} we focus on the process of local cascade which represents an interesting example of persistence of inhomogeneity at small scales. Section~\ref{sec:fseac} deals with the interplay between inhomogeneity, anisotropy and finite-size effects, which provides a correction to the dominant isotropic behaviour. Conclusions and possible future developments follow in \S~\ref{sec:conc}. The appendix is devoted to the reformulation of the problem in a finite box in the presence of periodic boundary conditions.

\section{Basic equations} \label{sec:be}

\subsection{Kraichnan advection model}

 Let us consider a passive scalar field $\theta(\bm{x},t)$ transported by a turbulent flow:
 \begin{equation} \label{csp}
  \partial_t\theta+\bm{v}\cdot\bm{\partial}\theta=\kappa\partial^2\theta+f\;,
 \end{equation}
 where $\kappa$ is the molecular diffusivity. The incompressible velocity field $\bm{v}(\bm{x},t)$ is assumed statistically homogeneous and isotropic, whereas the source term $f(\bm{x},t)$ is allowed not to be invariant under translations: a relevant example is provided by the emission of a tracer from a point source, located e.g.~in the origin.\\
 Let us now specialize to the Kraichnan ensemble \cite[][]{K68,K94}, where the velocity is a Gaussian, zero-average, white-in-time field with two-point correlation whose spatial behaviour is described by
 \begin{equation} \label{fcv}
  D_{\mu\nu}(\bm{x}_1,\bm{x}_2)=D_0\delta_{\mu\nu}-d_{\mu\nu}(\bm{x}_1-\bm{x}_2)\;.
 \end{equation}
 The second-order moment of the velocity increments is given by
 \begin{equation} \label{fsv}
  d_{\mu\nu}(\bm{r})=D_1r^{\xi}\left[(d+\xi-1)\delta_{\mu\nu}-\xi\frac{r_{\mu}r_{\nu}}{r^2}\right]
 \end{equation}
 for $r=|\bm{r}|$ smaller than the integral scale of the velocity field ($L_v$), above which $d_{\mu\nu}(\bm{r})$ saturates to an almost constant value whose order of magnitude is $D_1L_v^{\xi}$. Consequently, since the correlation $D_{\mu\nu}(\bm{r})$ has to vanish for $r\to\infty$, the relation $D_0\sim D_1L_v^{\xi}$ holds. Here, $d$ is the space dimension ($\ge 2$) and $\xi$ is the scaling exponent, describing the degree of roughness present in the velocity field, lying in the interval $(0,2)$.

 The two-point equal-time correlation function $C(\bm{x}_1,\bm{x}_2,t)=\langle\theta(\bm{x}_1,t)\theta(\bm{x}_2,t)\rangle$ may be expressed as a function of the center of mass $\bm{z}=(\bm{x}_1+\bm{x}_2)/2$ and of the separation $\bm{r}=\bm{x}_1-\bm{x}_2$. In these coordinates the equation for the correlation function $C(\bm{r},\bm{z},t)$ follows from the application of Gaussian integration by parts \cite[][]{F63,N65,D64}. Its form reads
 \begin{equation} \label{c2}
  \partial_tC=[2\kappa\delta_{\mu\nu}+d_{\mu\nu}(\bm{r})]\frac{\partial^2 C}{\partial r_{\mu}\partial r_{\nu}}+\frac{(D_0+2\kappa)\delta_{\mu\nu}+D_{\mu\nu}(\bm{r})}{4}\frac{\partial^2 C}{\partial z_{\mu}\partial z_{\nu}}+F\;,
 \end{equation}
 where $F(\bm{r},\bm{z})$ represents the correlator $\langle\theta(\bm{x}_1,t)f(\bm{x}_2,t)+\theta(\bm{x}_2,t)f(\bm{x}_1,t)\rangle$.\\
 Two cases appear to be quite relevant, also in connection with applications: the case of a constant emission from a point source and the one where the emission is random in time (but still punctual in space). The former case turns out to be quite cumbersome to be attacked by analytical methods and is still under investigation. Here, we shall focus on a Gaussian, zero-average, white-in-time forcing representing a random emission from the origin. Namely, $f(\bm{x},t)=f_0(t)\delta(\bm{x})$ with $\langle f(\bm{x}_1,t_1)f(\bm{x}_2,t_2)\rangle=F_0\delta(\bm{x}_1)\delta(\bm{x}_2)\delta(t_1-t_2)$, so that $F(\bm{r},\bm{z})=F_0\delta(\bm{r})\delta(\bm{z})$. As we shall see in detail, this case is amenable to analytical treatment.

\subsection{Fourier transform and $\mathrm{SO}(d)$ decomposition}

 Let us now come back to (\ref{c2}). Fourier transforming it in $\bm{z}$ and defining
 \[\hat{C}(\bm{r},\bm{q},t)=\!\int\!\ud^d\bm{z}\,\mathrm{e}^{-\mathrm{i}\bm{q}\cdot\bm{z}}C(\bm{r},\bm{z},t)\;,\qquad\hat{F}(\bm{r},\bm{q})=\!\int\!\ud^d\bm{z}\,\mathrm{e}^{-\mathrm{i}\bm{q}\cdot\bm{z}}F(\bm{r},\bm{z})\;,\]
 we obtain:
 \begin{equation} \label{errequ}
  \partial_t\hat{C}=\left[2\kappa\delta_{\mu\nu}+d_{\mu\nu}(\bm{r})\right]\frac{\partial^2\hat{C}}{\partial r_{\mu}\partial r_{\nu}}-\frac{(D_0+2\kappa)\delta_{\mu\nu}+D_{\mu\nu}(\bm{r})}{4}q_{\mu}q_{\nu}\hat{C}+\hat{F}\;.
 \end{equation}
 In the present case, the forcing-correlation transformed $\hat{F}=F_0\delta(\bm{r})$ is independent of the wavenumber.\\
 Equation (\ref{errequ}) is differential only in $\bm{r}$ and is algebraic in the centre-of-mass wavenumber $\bm{q}$. The second term on the right-hand side represents the inhomogeneous contribution and consistently vanishes for $q=0$, which is equivalent to average all over the space. It is convenient to rewrite its $\bm{r}$-dependent coefficient in the following way:
 \begin{eqnarray} \label{coeffic}
  \displaystyle-\frac{(D_0+2\kappa)\delta_{\mu\nu}+D_{\mu\nu}(\bm{r})}{4}&=&\displaystyle-\left[\frac{D_0+\kappa}{2}-\frac{(d-1)(d+\xi)}{4d}D_1r^{\xi}\right]\delta_{\mu\nu}\nonumber\\
  &&\displaystyle+\frac{\xi}{4d}D_1r^{\xi}\left(\delta_{\mu\nu}-d\frac{r_{\mu}r_{\nu}}{r^2}\right)\;.
 \end{eqnarray}
 Substituting it back, it is clear that the last term in (\ref{coeffic}) generates the only contribution in (\ref{errequ}) not invariant under rotations of $\bm{r}$, because it gives rise to a scalar product between $\bm{r}$ and $\bm{q}$ that mixes different angular sectors. However, at separations $r\ll L_v$, a simplification is possible, since, in that case, the order of magnitude of $d_{\mu\nu}(\bm{r})\approx D_1r^{\xi}$ is negligible with respect to $D_0\sim D_1L_v^{\xi}$. Therefore, $D_{\mu\nu}(\bm{r})\simeq D_0\delta_{\mu\nu}$ and the right-hand side of (\ref{coeffic}) simplifies into $-(D_0+\kappa)\delta_{\mu\nu}/2$. It is worth noticing that, when $r$ is of the order of (or larger than) $L_v$, a coupling between anisotropy and inhomogeneity takes place: we shall come back to this point in \S~\ref{sec:fseac}, where the consequences of keeping the full form (\ref{coeffic}) into account will be discussed. Here, we concentrate on the sole case $r\ll L_v$ in the stationary state with vanishing diffusivity\footnote{Attention should, in principle, be paid to the limit of vanishing diffusive scale, but for the rough flows ($\xi\ne2$) considered here no commutation problem arises with the limit of vanishing forcing correlation length $L$.}
 and we can thus consider the simpler equation
 \begin{equation} \label{c2t}
  d_{\mu\nu}(\bm{r})\frac{\partial^2\hat{C}}{\partial r_{\mu}\partial r_{\nu}}-\frac{1}{2}D_0q^2\hat{C}+\hat{F}=0\;.
 \end{equation}
 A dimensional-analysis balance between the first and the second term in (\ref{c2t}) leads to the introduction of a new scale
 \[\ell_q=\left[\frac{q^2D_0}{2(d-1)D_1}\right]^{-1/(2-\xi)}\;,\]
 which is associated to the strength of the scalar inhomogeneities and measures the separation above which they become relevant. A decomposition on the spherical harmonics \cite[][]{BP05},
 \[\hat{C}(\bm{r},\bm{q})=\sqrt{\Omega}\sum_{l,m}\hat{C}_{l,m}(r,\bm{q})Y_{l,m}(\Phi)\;,\qquad\hat{F}(\bm{r},\bm{q})=\sqrt{\Omega}\sum_{l,m}\hat{F}_{l,m}(r,\bm{q})Y_{l,m}(\Phi)\;,\]
 with $\Phi$ denoting the solid angle associated with $\bm{r}$ and $\Omega$ its overall value, yields the following equation for $\hat{C}_{l,m}(r,\bm{q})$ in each sector:
 \begin{equation} \label{eq:1}
  r^{-(d-1)}\partial_rr^{d+\xi-1}\partial_r\hat{C}_l-\frac{(d+\xi-1)l(d-2+l)}{d-1}r^{-2}\hat{C}_l-\ell_q^{-(2-\xi)}\hat{C}_l+\varphi_l=0\;.
 \end{equation}
 Note that, because of foliation on $l$ and degeneration, we have dropped the dependence on the subscript $m$ and we have introduced the rescaled forcing $\varphi_l(r)=\hat{F}_l(r,\bm{q})/(d-1)D_1$ (independent of $\bm{q}$ because of the punctual nature of the source).

\section{General solution} \label{sec:gs}

 The general solution of (\ref{eq:1}), as a function of $r$ and $\ell_q$, reduces to the zero mode \cite[][]{MAS05}
 \begin{equation} \label{gszm}
  \hat{C}_l(r;\ell_q)=w^{-\nu_0}\left[A_lK_{\nu_l}(w)+B_lI_{\nu_l}(w)\right]\;,
 \end{equation}
 where $w=2(2-\xi)^{-1}(r/\ell_q)^{(2-\xi)/2}$ and $\nu_l=[(d+\xi-2)^2+4(d+\xi-1)l(d-2+l)/(d-1)]^{1/2}/(2-\xi)$.\\
 To determine the coefficients $A_l$ and $B_l$, one can approximate the Dirac $\delta$ by a Heaviside $\Theta$, exploiting the vanishing of $\varphi_l$ in all the anisotropic sectors $l\ne0$:
 \[\delta(\bm{r})=\frac{r^{-(d-1)}}{\Omega}\delta(r)=\lim_{L\to0}\frac{r^{-(d-1)}}{\Omega L}\Theta(L-r)\Rightarrow\varphi_0(r)=\lim_{L\to0}\frac{F_0r^{-(d-1)}}{(d-1)D_1\Omega L}\Theta(L-r)\;.\]
 It must be stressed that, in this way, the point source is obtained as a limit of a forcing having positive Corrsin integral $Q_0\equiv\!\int\!\ud^d\bm{r}\,\hat{F}(\bm{r},\bm{q})=F_0$. The investigation of the case $Q_0=0$ \cite[][]{FF05,CS05,CS06} will be left for future investigation.\\
 Studying the solution also for $r<L$, matching the solution $\hat{C}_l$ and its first derivative in $r=L$, imposing regularity for small $r$ and vanishing for large $r$ \cite[][]{MAS05}, and eventually taking the limit $L\to0$, one finds $B_l=0\;\forall l$ and
 \[A_l=\left(\frac{2-\xi}{2}\right)^{\nu_0}\ell_q^{(d-\xi+2)/2}\lim_{L\to0}\!\int_0^W\!\ud\omega\,\varphi_l(\rho)\omega^{\nu_0+1}I_{\nu_l}(\omega)=\delta_{l,0}k_{\dag}\frac{F_0}{D_1}\ell_q^{2-d-\xi}\;,\]
 where $\omega\equiv w|_{r=\rho}=2(2-\xi)^{-1}(\rho/\ell_q)^{(2-\xi)/2}$, $W\equiv w|_{r=L}=2(2-\xi)^{-1}(L/\ell_q)^{(2-\xi)/2}$ and $k_{\dag}=2(2-\xi)^{-d/(2-\xi)}/(d-1)\Omega\Gamma(\nu_0+1)$ ($\Gamma(\cdot)$ being Euler's function).\\
 In the pseudospectral space $(\bm{r},\bm{q})$ the scalar-correlation transformed thus coincides with its isotropic projection and depends only on the moduli $r$ and $q$ (i.e.~$\ell_q$) as
 \[\hat{C}(r;\ell_q)=k_{\dag}\frac{F_0}{D_1}\ell_q^{-(d+\xi-2)/2}r^{-(d+\xi-2)/2}K_{\nu_0}(w)\;.\]
 Back to the physical space, the correlation is thus independent of the angle between $\bm{r}$ and $\bm{z}$ and is a function of $r$ and $z$ only:
 \begin{equation} \label{sol}
  C(r,z)=k_{\ddag}\frac{F_0}{D_1}\left(\frac{D_1}{D_0}\right)^{d/2}r^{d\xi/2-2d-\xi+2}\left[1+\frac{\Omega(2-\xi)^2}{4\upi}\frac{D_1}{D_0}z^2r^{-(2-\xi)}\right]^{-\frac{\scriptstyle d(4-\xi)}{\scriptstyle2(2-\xi)}+1}\;.
 \end{equation}
 For $d=2$, $k_{\ddag}=k_{\dag}2^{-2}\upi^{-1}(2-\xi)^{(4-\xi)/(2-\xi)}\Gamma[2/(2-\xi)]$; for $d=3$, $k_{\ddag}=k_{\dag}2^{-1}\upi^{-3/2}(2-\xi)^{(7-2\xi)/(2-\xi)}\Gamma[3/2+(1+\xi)/(2-\xi)]$.\\
 It is worth noticing that the behaviour $C\sim r^{-(d+\xi-2)}$, which is typical of the homogeneous situation \cite[][]{FGV01} for $r\gg L$, is not observed in this case (even if $L=0$), unless one integrates the correlation on the whole space, thus averaging out the inhomogeneity and defining the contribution in the homogeneous `sector'. In the pseudo-spectral space, this last operation is equivalent to consider $q=0$ ($\Rightarrow\ell_q\to\infty\Rightarrow w=0$) and thus corresponds to keep into account only the leading term in the development of $K_{\nu_0}(w)$ for small arguments.

\section{Local cascade and small-scale persistence of inhomogeneity} \label{sec:lc}

\begin{figure}
  \includegraphics[height=8cm]{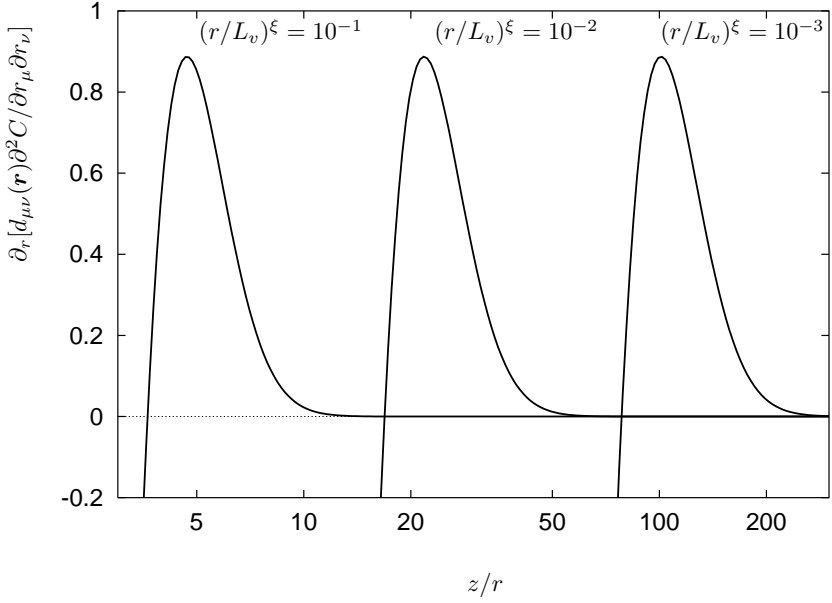}
  \caption{Derivative of the left-hand side of equation (\ref{cfps}) respect to $r$, plotted \textsl{vs} $z/r$ for $\xi=4/3$ and $d=3$. It is evident how the ratio $r/L_v$, labelling the three curves, affects the limits of the range in which approximation (\ref{ultima}) is valid and a constant flux holds.}
  \label{errezeta}
 \end{figure}

 Recalling that the ratio $D_1/D_0$ appearing in (\ref{sol}) is of the order of $L_v^{-\xi}$, two opposite developments are meaningful, corresponding to small or large values of the quantity
 \begin{equation} \label{esse}
  s\equiv\left(\frac{z}{r}\right)^2\left(\frac{r}{L_v}\right)^{\xi}\;.
 \end{equation}
 This defines a new characteristic length scale,
 \begin{equation} \label{lz}
  \mathcal{L}_z\equiv z^{2/(2-\xi)}L_v^{-\xi/(2-\xi)}\;,
 \end{equation}
 dependent (monotonically) on $z$ and whose meaning as the local integral scale of scalar turbulence will be clear shortly.\\
 For small $s$ (i.e.~$r\gg\mathcal{L}_z$) the correlation is approximated by a power law in $r$, $C\sim L_v^{-d\xi/2}r^{-d(4-\xi)/2+2-\xi}$. On the contrary, for large $s$ (i.e.~$r\ll\mathcal{L}_z$), a power law in $z$ is found, and $r$ appears only in subleading terms:
 \begin{equation} \label{ultima}
  C(r,z)\sim\langle\theta^2(z)\rangle-\textrm{const.}\times\epsilon(z)r^{2-\xi}\;,
 \end{equation}
 Here, $\langle\theta^2(z)\rangle\equiv(F_0/D_1)L_v^{\xi(d+\xi-2)/(2-\xi)}z^{2-d(4-\xi)/(2-\xi)}$ and $\epsilon(z)\equiv\langle\theta^2(z)\rangle/\mathcal{L}_z^{2-\xi}$ represent the scalar variance and the local dissipation, respectively.\\
 Expression (\ref{ultima}) proves that, at scales smaller than $\mathcal{L}_z$ (i.e.~for $r$ sufficiently smaller than $z$, according to the value of $L_v$), the increment of the two-point equal-time correlation function has the same behaviour shown by the homogeneous case (power law in $r$ with exponent $2-\xi$: see e.g.~Falkovich \emph{et al.} 2001) at scales smaller than the forcing integral scale $L$. This result suggests that a cascade-like mechanism might thus be present also in this case, even if here $r>L$ by construction.\\
 This possibility can actually be supported by the following considerations. The velocity field sweeps the scalar, initially concentrated where it was released, and generates structures which, $\forall\bm{x}_1$, are correlated on the scale $\bm{x}_1$. This amounts to saying that correlations between each point $\bm{x}_1$ and the origin $\bm{x}_2=\bm{0}$ are created. In the centre-of-mass frame of reference, this means that in every point $\bm{z}$ ($=\bm{x}_1/2$) a local cascade can then take place, starting from separations $r$ sufficiently smaller than $z$.\\
 The quantity $\mathcal{L}_z$ thus plays the role of an effective local forcing correlation length: the word `local' here refers to the fact that, while in the homogeneous situation the prefactors are expressed in terms of constant quantities, on the contrary in this case a dependence of $\langle\theta^2(z)\rangle$ and $\epsilon(z)$ on the point still persists. It is also worth noticing from (\ref{lz}) that, for very rough flows ($\xi\to0$), $\mathcal{L}_z$ becomes proportional to $z$, consistently with the impossibility of a consistent definition of $L_v$. On the contrary, for almost smooth flows ($\xi\to2$), one finds $\mathcal{L}_z\sim(z/L_v)^{2/(2-\xi)}$, critically dependent on whether the center-of-mass distance from the source lies within the velocity correlation range.\\
 The aforementioned physical interpretation of local cascade can easily be supported mathematically by considering the physical-space counterpart of (\ref{c2t}), i.e.~the simplified form of (\ref{c2}) with the usual approximations $r\ll L_v$ and $\kappa=0=\partial_t$:
 \begin{equation} \label{2de}
  d_{\mu\nu}(\bm{r})\frac{\partial^2C}{\partial r_{\mu}\partial r_{\nu}}+\frac{D_0}{2}\frac{\partial^2C}{\partial z_{\mu}\partial z_{\mu}}+F=0\;.
 \end{equation}
 In the homogeneous case, the absence of any dependence on $\bm{z}$ gives rise to the convective-range balance
 \begin{equation} \label{cfh}
  d_{\mu\nu}(\bm{r})\frac{\partial^2C}{\partial r_{\mu}\partial r_{\nu}}=-F(r)\,:
 \end{equation}
 the derivative with respect to $r$ of the left-hand side of (\ref{cfh}) vanishes in the presence of a constant corresponding right-hand side, as is often the case. On the other hand, with point-source forcing, away from the origin, the balance (\ref{2de}) has to be written as
 \begin{equation} \label{cfps}
  d_{\mu\nu}(\bm{r})\frac{\partial^2C}{\partial r_{\mu}\partial r_{\nu}}=-\frac{D_0}{2}\frac{\partial^2C}{\partial z_{\mu}\partial z_{\mu}}\;,
 \end{equation}
 but the vanishing of the derivative of the left-hand side of (\ref{cfps}) still takes place for $r$ sufficiently smaller than $z$. This is shown in figure \ref{errezeta},
 which moreover reflects how this interpretation has its validity limit affected by a change in the ratio $r/L_v$ appearing in the adimensional parameter $s$ (\ref{esse}). One should indeed remember that the three scales $r$, $z$ and $L_v$ appear in a nontrivial way in $s$, whose magnitude is the key point for approximation (\ref{ultima}) and its consequences.\\
 The present case thus represents an interesting example of a situation in which inhomogeneity persists at small scales, as the behaviour $C\sim r^{2-\xi}$ observed at small $r$ does not correspond to what was defined at the end of \S~\ref{sec:gs} to be interpreted as the homogeneous counterpart ($C\sim r^{-(d+\xi-2)}$).

\section{Finite-size effects and anisotropic contributions} \label{sec:fseac}

 A comment is required about the relevance of the so-called \emph{finite-size effects}. In other words, one would like to quantify the error deriving from the approximation $r\ll L_v$, which was used to simplify (\ref{coeffic}) and thus to decouple inhomogeneity from anisotropy. This quantification becomes possible, if one proceeds in the following way. First of all, one should notice that, after the decomposition into spherical harmonics, no more foliation takes place. Namely, the equation for the isotropic sector is still a closed one (with the appearance of a new term, according to the contribution within square braces in (\ref{coeffic})),
 \[r^{-(d-1)}\partial_r\left(r^{d+\xi-1}\partial_r\hat{C}_0\right)-\ell_q^{-(2-\xi)}\left[1-\frac{(d+\xi)D_1}{2d(d-1)D_0}r^{\xi}\right]\hat{C}_0+\varphi_0(r)=0\;,\]
 and gives
 \begin{equation} \label{csi}
  \hat{C}_0\propto r^{-(d+\xi-2)/2}K_{\nu_0}(w)\left[1+O\left(\frac{r}{L_v}\right)^{\xi}\right]\;,
 \end{equation}
 but $\hat{C}_0$ now enters the equation for $l=2$ as a forcing term (the $l=1$ sector remains unforced because this procedure only couples even sectors, as can be deduced by decomposition (\ref{coeffic})). A simple power-counting operation is possible in Fourier space for $r\ll\ell_q$, where (specifying the order of the error in the development of $K_{\nu_0}(w)$ for small arguments)
 \begin{equation} \label{duemenocsi}
  \hat{C}_0\propto r^{-(d+\xi-2)}\left[1+O\left(\frac{r}{\ell_q}\right)^{2-\xi}\right]\;.
 \end{equation}
 As a result, in this regime one easily obtains
 \begin{equation} \label{l2}
  \hat{C}_2\sim L_v^{-\xi}\ell_q^{-(2-\xi)}r^{-(d+\xi-4)}\sim\left(\frac{r}{L_v}\right)^{\xi}\left(\frac{r}{\ell_q}\right)^{2-\xi}\hat{C}_0\;.
 \end{equation}
 Equation (\ref{l2}) shows that the first excited anisotropic sector carries a factor, with respect to the isotropic solution, given by the product between the corrections in (\ref{csi}) and in (\ref{duemenocsi}). Its interpretation is thus very simple and meaningful: at the lowest order, the most relevant anisotropic correction derives from the coupling of finite-size effects ($O(r/L_v)^{\xi}$) and of inhomogeneities ($O(r/\ell_q)^{2-\xi}$).\\
 It can also be shown that, in the opposite situation ($r\gg\ell_q$), $\hat{C}_2$ is still given by the right-hand side of (\ref{l2}) but without the factor $(r/\ell_q)^{2-\xi}$. This implies that, back to the physical space, the leading contribution is always the isotropic one, provided that $r\ll L_v$. The higher-$l$ anisotropic terms $\hat{C}_4$, $\hat{C}_6$, etc.~are indeed smaller and smaller, because they are forced by the (small) quantities $\hat{C}_2$, $\hat{C}_4$, etc.~respectively.\\
 Such anisotropic corrections are expected to play a non-negligible role only when the scales $r$ and $z$ are comparable, but not when either is much greater than the other. An example of the former case is provided, for $z=r/2$, by the comparison between the situations $\bm{z}\parallel\bm{r}$ (where one of the two points in which the correlation is calculated lies on the source) and $\bm{z}\perp\bm{r}$ (where both points are $\sqrt{2}z$ away from the origin): a difference must clearly exist, but cannot be caught by the isotropic function $C_0(r,z)$ and turns out to be subdominant. On the contrary, if $r\gg z$ the two points are almost symmetric with respect to the origin, and if $r\ll z$ their relative separation is much smaller than their distance from the source: in both cases, a rotation of $\bm{r}$ with fixed $\bm{z}$ would change little. Of course, the problem is always invariant under rigid rotations of the whole space (and thus of both vectors $\bm{r}$ and $\bm{z}$) around the origin.

\section{Conclusions and perspectives} \label{sec:conc}

 The dynamics of a passive scalar released from a point source has been investigated in this paper, as a prototype of inhomogeneity. Focusing on the Kraichnan advection model with Gaussian, white-in-time and zero-mean forcing, it has been possible to study analytically the two-point equal-time scalar correlation function and to prove a persistence of inhomogeneity at small scales, in the spirit of the local cascade process described in \S~\ref{sec:lc}. Still to be understood in more detail is the interplay between inhomogeneity, anisotropy and finite-size effects mentioned in \S~\ref{sec:fseac}.\\
 An interesting open problem is the extension of our calculation to higher-order scalar correlation functions, with the aim of corroborating our results obtained for the two-point correlation function.\\
 Another natural extension of the present work would be represented by the study of a point source not satisfying the aforementioned hypotheses, e.g.~a constant-in-time scalar emission. This situation is already under investigation, and a fully-analytical study should be completed by numerical results. A related issue is understanding how these results should change in the presence of smooth flows ($\xi=2$).\\
 Inhomogeneities were here limited to the forcing term, while the velocity field was assumed homogeneous and isotropic. It would be of interest to reformulate the problem with velocity ensembles not invariant under translation, so as to deal with a completely inhomogeneous situation. Moreover, whenever the Kraichnan model is used, the question of a possible extension to more realistic flows arises: numerical simulations would thus be required to test the present results with actual solutions of the Navier--Stokes equation.\\
 Lastly, a coarse-grained description might be applied to the present case, with the aim of finding large-scale closed equations and an explicit exact parameterization for the small scales, accounting for the effects played by inhomogeneities. The infrared limit of the theory for the passive scalar has already been introduced with good results in the homogeneous isotropic case \cite[][]{MACFM03,MACM04,CMAM05,CMAM06}. The extension to inhomogeneous cases is in order, in particular using the point-source problem as a paradigm.

\begin{acknowledgements}
 AM and MMA have been partially supported by COFIN 2005 project n.~2005027808 and by CINFAI consortium.  Part of this work has been done within the 2006 CNR Short-Term Mobility programme (AM). MMA acknowledges useful discussions with Grisha Falkovich and Stefano Musacchio.
\end{acknowledgements}

\oneappendix

\section{Periodicity and discrete spectrum}

 It is interesting to reformulate the point-source problem in a finite $d$-dimensional box of side $a$ with periodic boundary conditions, which is equivalent to consider an infinite $d$-dimensional grid of point sources with mesh size $a$. It can easily be shown that, upon Fourier transforming, the forcing-correlation spectrum is no longer continuum and flat, but is active only on a discrete set of wavenumbers ($\bm{q}_{\bm{k}}=2\upi\bm{k}/a,\ \forall\bm{k}\in\mathbb{Z}^d$) with uniform intensities. In order to reconstruct the correlation function $C(\bm{r},\bm{z})$ it is thus sufficient to analyse the discrete values $\hat{C}(\bm{r},\bm{q}_{\bm{k}})$ (we will only focus on the isotropic sector $l=0$).\\
 For $\bm{k}=\bm{0}=\bm{q}_{\bm{k}}$, the pure homogeneous scaling behaviour $\hat{C}\sim r^{-(d+\xi-2)}$ is obviously found. For $\bm{k}\neq\bm{0}$, such power law is replaced by Bessel functions, which can in turn be expanded in Taylor series, with a result analogous to (\ref{duemenocsi}). That is, for each wavenumber the leading behaviour for small $r$ is always given by the same homogeneous contribution, but for a fixed $r$ the corrections to such term become more and more relevant with growing modulus of $\bm{q}_{\bm{k}}$. When antitransforming, one would be tempted by extrapolating such leading behaviour from each mode and concluding that $C\sim r^{-(d+\xi-2)}$ for small $r$ also upon superposition. This is clearly not the case, as the behaviour $C\sim r^{2-\xi}$ is actually found for small $r$. In other words, it is not possible to exchange the Taylor and Fourier series, corresponding to the power expansion of the Bessel function and to the discrete antitransform respectively, because of the absence of uniform convergence.\\
 It is worth noticing that, in this consideration, we did not assume any power-law behaviour for each mode (differently from what is usually done in each anisotropic sector upon $\mathrm{SO}(d)$ decomposition), coherently with the result that for each wavenumber we obtain a Bessel function, i.e.~an infinite superposition of power laws.


\begin{thebibliography}{99}
 \bibitem[Basu, Foufoula-Georgiou \& Port\'e-Agel(2004)]{BFP04}
  {\sc Basu, S., Foufoula-Georgiou, E. \& Port\'e-Agel, F.} 2004
  Synthetic turbulence, fractal interpolation and large-eddy simulation.
  {\it Phys. Rev.} E {\bf 70}, 026310.
 \bibitem[Biferale \& Procaccia(2005)]{BP05}
  {\sc Biferale, L. \& Procaccia, I.} 2005
  Anisotropy in turbulent flows and in turbulent transport.
  {\it Phys. Rep.} {\bf 414}, 43--164.
 \bibitem[Celani, Martins Afonso \& Mazzino(2005)]{CMAM05}
  {\sc Celani, A., Martins Afonso, M. \& Mazzino, A.} 2005
  Coarse-grained scalar transport: closures and large-eddy simulations.
  In {\it Progress in Turbulence 2, Proceedings of the iTi Conference on Turbulence,
  Bad Zwischenahn (Germany), September 25--28, 2005}
  (ed.~M.~Oberlack, S.~Guenther, T.~Weller, G.~Khujadze, A.~Osman, M.~Frewer \& J.~Peinke), Springer (in press).
 \bibitem[Celani, Martins Afonso \& Mazzino(2006)]{CMAM06}
  {\sc Celani, A., Martins Afonso, M. \& Mazzino, A.} 2006
  Coarse-grained description of a passive scalar.
  {\it J. Turb.} {\bf 7} (52), 1--18.
 \bibitem[Celani \& Seminara(2005)]{CS05}
  {\sc Celani, A. \& Seminara, A.} 2005
  Large-Scale Structure of Passive Scalar Turbulence.
  {\it Phys. Rev. Lett.} {\bf 94}, 214503.
 \bibitem[Celani \& Seminara(2006)]{CS06}
  {\sc Celani, A. \& Seminara, A.} 2006
  Large-Scale Anisotropy in Scalar Turbulence.
  {\it Phys. Rev. Lett.} {\bf 96}, 184501.
 \bibitem[Donsker(1964)]{D64}
  {\sc Donsker, M.~D.} 1964
  On function space integrals.
  In {\it Proceedings of a Conference on the Theory and Applications of Analysis in Function Space,
  Dedham (MA), June 9--13, 1963}
  (ed.~W.T.~Martin \& I.~Segal),
  ch.~2, pp.~17--30, MIT Press.
 \bibitem[Falkovich \& Fouxon(2005)]{FF05}
  {\sc Falkovich, G. \& Fouxon, A.} 2005
  Anomalous Scaling of a Passive Scalar in Turbulence and in Equilibrium.
  {\it Phys. Rev. Lett.} {\bf 94}, 214502.
 \bibitem[Falkovich, Gaw\c{e}dzki \& Vergassola(2001)]{FGV01}
  {\sc Falkovich, G., Gaw\c{e}dzki, K. \& Vergassola, M.} 2001
  Particles and fields in fluid turbulence.
  {\it Rev. Mod. Phys.} {\bf 73}, 913--975.
 \bibitem[Furutsu(1963)]{F63}
  {\sc Furutsu, K.} 1963
  On the statistical theory of electromagnetic waves in a fluctuating medium.
  {\it J. Res. Nat. Bur. Standards} D {\bf 67}, 303--323.
 \bibitem[Kraichnan(1968)]{K68}
  {\sc Kraichnan, R.~H.} 1968
  Small-scale structure of a scalar field convected by turbulence.
  {\it Phys. Fluids} {\bf 11}, 945--953.
 \bibitem[Kraichnan(1994)]{K94}
  {\sc Kraichnan, R.~H.} 1994
  Anomalous scaling of a randomly advected passive scalar.
  {\it Phys. Rev. Lett.} {\bf 72}, 1016--1019.
  \bibitem[Martins Afonso \& Sbragaglia(2005)]{MAS05}
  {\sc Martins Afonso, M. \& Sbragaglia, M.} 2005
  Inhomogeneous Anisotropic Passive Scalars.
  {\it J. Turb.} {\bf 6} (10), 1--13.
 \bibitem[Martins Afonso \emph{et al.}(2003)]{MACFM03}
  {\sc Martins Afonso, M., Celani, A., Festa, R. \& Mazzino, A.} 2003
  Large-eddy-simulation closures of passive scalar turbulence: a systematic approach.
  {\it J. Fluid Mech.} {\bf 496}, 355--364.
 \bibitem[Martins Afonso, Celani \& Mazzino(2004)]{MACM04}
  {\sc Martins Afonso, M., Celani, A. \& Mazzino, A.} 2004
  Closures for large-eddy simulations of passive scalars.
  In {\it Advances in Turbulence X, Proceedings of the Tenth European Turbulence Conference,
  Trondheim (Norway), June 29--July 2, 2004}
  (ed.~H.I.~Andersson \& P.A.~Krogstad), pp.~319--322, CIMNE.
 \bibitem[Novikov(1965)]{N65}
  {\sc Novikov, E.~A.} 1965
  Functionals and the random-force method in turbulence theory.
  {\it Sov. Phys.} JETP {\bf 20}, 1290--1294.
 \bibitem[Scotti \& Meneveau(1997)]{SM97}
  {\sc Scotti, A. \& Meneveau, C.} 1997
  Fractal model for coarse-grained partial differential equations.
  {\it J. Phys. Rev. Lett.} {\bf 78}, 867--870.
 \bibitem[Toschi \emph{et al.}(1999)]{TASBP99}
  {\sc Toschi, F., Amati, G., Succi, S., Benzi, R. \& Piva, R.} 1999
  Intermittency and Structure Functions in Channel Flow Turbulence.
  {\it Phys. Rev. Lett.} {\bf 82}, 5044--5047.
\end{thebibliography}
\end{document}